\documentclass[12pt]{article}
\pdfoutput=1
\usepackage{epsf,amsfonts,amssymb,epsfig,amsmath,mathtools,graphics,slashed}
\usepackage{hyperref,hep,graphicx,subfig}

\makeatletter

\newcommand{\Rmnum}[1]{\expandafter\@slowromancap\romannumeral #1@}
\makeatother

\newcommand{\nn}{\notag \\}

\newcommand{\hsce}{\zeta}

\begin{document}

\makeatletter
\renewcommand{\theequation}{\thesection.\arabic{equation}}
\@addtoreset{equation}{section}
\makeatother

\baselineskip 18pt

\begin{titlepage}

\vfill

\begin{flushright}
Imperial/TP/2014/JG/03\\
\end{flushright}

\vfill

\begin{center}
   \baselineskip=16pt
   {\Large\bf Thermoelectric DC conductivities\\from black hole horizons}
  \vskip 1.5cm
  \vskip 1.5cm
      Aristomenis Donos$^1$ and Jerome P. Gauntlett$^2$\\
   \vskip .6cm
    \vskip .6cm
      \begin{small}
      \textit{$^1$DAMTP, 
       University of Cambridge\\ Cambridge, CB3 0WA, U.K.}
        \end{small}\\
      \vskip .6cm
      \begin{small}
      \textit{$^2$Blackett Laboratory, 
        Imperial College\\ London, SW7 2AZ, U.K.}
        \end{small}\\*[.6cm]

\end{center}

\vfill

\begin{center}
\textbf{Abstract}
\end{center}

\begin{quote}
An analytic expression for the DC electrical conductivity in terms of black hole horizon data was recently
obtained for a class of holographic black holes exhibiting momentum dissipation. We generalise this result to obtain analogous expressions 
for the DC thermoelectric and thermal conductivities. We illustrate our results 
using some holographic Q-lattice black holes as well as for some
black holes with linear massless axions, in both $D=4$ and $D=5$ bulk spacetime dimensions, which include both spatially isotropic and
anisotropic examples.
We show that some recently constructed ground states of holographic Q-lattices, which can be either electrically insulating or metallic,
are all thermal insulators.
\end{quote}

\vfill

\end{titlepage}
\setcounter{equation}{0}


\section{Introduction}

A striking feature of holography is that it provides a prescription for calculating
transport coefficients of strongly coupled systems 
by analysing small perturbations about the black holes that describe the equilibrium state. 
In this paper we will be interested in strongly coupled CFTs at finite temperature, $T$, and chemical potential, $\mu$,
with respect to an abelian global symmetry,
which are described by asymptotically AdS black holes which are electrically charged if $\mu\ne 0$. The addition of a small
electric field $E_i$ and/or thermal gradient $\nabla_i T$ will induce an electric current $J^i$ and a heat current $Q^i=T^{ti}-\mu J^i$, where
$T^{ab}$ is the stress tensor and
$i$ labels a spatial index. At linearised order we have the generalised Ohm/Fourier law:
\begin{align}\label{bigform}
\left(
\begin{array}{c}
J\\Q
\end{array}
\right)=
\left(\begin{array}{cc}
\sigma & \alpha T \\
\bar\alpha T &\bar\kappa T   \\
\end{array}\right)
\left(
\begin{array}{c}
E\\-(\nabla T)/T
\end{array}
\right)\,.
\end{align}
In this expression the matrix $\sigma$ is the electrical conductivity, $\alpha$, $\bar\alpha$ are the thermoelectric conductivities 
and $\bar \kappa$ is the thermal conductivity.

Much work has been done on obtaining the AC conductivities, particularly $\sigma$, in a variety of different contexts, by allowing for 
perturbations with a time dependence of the form $e^{-i\omega t}$ and then imposing ingoing boundary conditions 
at the black hole event horizon. These calculations, reviewed in \cite{Hartnoll:2009sz,Herzog:2009xv}, 
utilise the fact that the conductivities are related to retarded Green's functions via $i\omega\sigma= G^R_{JJ}(\omega)$ etc. 
The DC conductivities, if they are finite, can then be obtained by carefully taking the $\omega\to 0$ limit.

Here we want to directly calculate the DC response. 
It has been known for some time that for the case for translationally invariant CFTs
with $\mu=0$ that are described by the AdS-Schwarzschild black brane solution \cite{Iqbal:2008by}, 
it is possible to obtain an expression for $\sigma_{DC}$ in terms of horizon data, somewhat analogous to the result
for the shear viscosity \cite{Policastro:2001yc,Kovtun:2004de}. However, when $\mu\ne 0$
a consideration of the standard AdS-RN black brane, describing a translationally invariant system, 
reveals that this is not always the case. Specifically, 
the translation invariance and finite charge density imply that the DC conductivities are infinite. 
More precisely, the real parts of the AC conductivities have a delta function at zero frequency.

Nonetheless, for holographic lattices associated with black holes with broken translation invariance, or more generally when
there is a mechanism for momentum dissipation, the associated 
DC conductivities will be finite
and one might hope to obtain a result in terms of black hole horizon data. This was recently confirmed for the electric conductivity $\sigma$ 
for a class of holographic Q-lattices in 
\cite{Donos:2014uba}\footnote{Similar results were obtained in \cite{Gouteraux:2014hca,Andrade:2013gsa}, using a different approach based on massive gravity, extending \cite{Davison:2013jba,Blake:2013bqa,Davison:2013txa}. A related result for inhomogeneous lattices, for small lattice strength, was obtained in \cite{Blake:2013owa}. Very recently the methods of \cite{Donos:2014uba} were used to obtain the electrical DC conductivity in the presence of a magnetic field in \cite{Blake:2014yla}.}. 
Such Q-lattices were introduced in \cite{Donos:2013eha} and break translation invariance periodically while preserving a homogeneous metric. 
This is a significant technical simplification since the black holes can be constructed by solving ODEs instead of PDEs\footnote{For the special case of $D=5$, helical lattices can also be constructed by solving ODEs \cite{Donos:2012js}, extending \cite{Iizuka:2012iv,Donos:2012gg}.
} as in 
the constructions \cite{Horowitz:2012ky,Horowitz:2012gs,Horowitz:2013jaa,Donos:2012js,Ling:2013nxa,Chesler:2013qla}.
It was shown in \cite{Donos:2014uba} that the electrical conductivity $\sigma$ can be expressed in terms of horizon data and 
here we generalise this analysis to obtain analogous expressions for $\alpha$,  $\bar \alpha$ and $\bar \kappa$. A key step in our derivation 
is to manipulate the bulk equations of motion to
obtain expressions for the electric and heat currents that are independent of the holographic radial coordinate and hence can
be evaluated at the black hole horizon.

For illustration we will mostly consider holographic Q-lattices in $D=4$ bulk dimensions which break translation invariance in either one or both of the two spatial directions of the dual CFT. 
In particular, they can be both isotropic and anisotropic. 
The matrices $\sigma,\alpha,\bar\kappa$ are diagonal and furthermore, since
the backgrounds are time-reversal invariant, we have a symmetric conductivity matrix with $\bar\alpha=\alpha$. 
It is straightforward to generalise our analysis to holographic Q-lattices in other dimensions
and indeed our final result can be cast in a $D$ independent manner. Our approach 
can also be extended to other holographic lattices, including inhomogeneous lattices \cite{Donos:2014yya}.

Our results give the DC conductivities for all temperatures. They also provide a convenient tool to extract the low-temperature
scaling behaviours of the DC conductivities when the far IR behaviour of the $T=0$ ground states are known. This allows
one to determine, for example, if the ground state is exhibiting
metallic or insulating behaviour with respect to the electric conductivity.
Specifically, one can construct
``small" black holes, by heating up the zero temperature 
IR ground states and then extract the low-temperature behaviour of the DC conductivity.

The low temperature scaling behaviour of the electrical conductivity was also discussed in \cite{Hartnoll:2012rj}
using the memory matrix formalism. The results here and in \cite{Donos:2014uba} complement and extend this illuminating work within a holographic setting. Specifically, the memory matrix formalism requires that the $T=0$ ground states are translationally invariant and that the strength of the lattice is small, whereas our approach has no such restrictions.

For the class of black holes that we consider we find, in general, that we can write
\begin{align}\label{ratintro}
\frac{\bar\kappa}{\alpha}=\frac{Ts}{q}\,,\qquad 
\bar L\equiv \frac{\bar\kappa}{\sigma T} \le \frac{s^2}{q^2}\,,
\end{align}
where $s$ and $q$ are the entropy density and electric charge of the black holes, respectively. 
In situations where the memory matrix formalism applies, using the results of \cite{Mahajan:2013cja}
we can rewrite these as
\begin{align}\label{newbdintro}
\frac{\bar\kappa}{\alpha}= \frac{\chi_{QP}}{\chi_{JP}}\,,\qquad \bar L \le \frac{1}{T^2}\frac{\chi_{QP}^2}{\chi_{JP}^2}\,,
\end{align}
where in the notation of \cite{Mahajan:2013cja},
$\chi$ are static susceptibilities involving the operators for the total momentum $P$, electric current $J$ and heat current $Q$.

We will illustrate our results for some known examples, including the isotropic and anisotropic Q-lattice black holes constructed in
\cite{Donos:2013eha,Donos:2014uba}. These include solutions where, at $T=0$, the black holes approach $AdS_2\times\mathbb{R}^{D-2}$ in the far IR, perturbed by irrelevant operators. We obtain low-temperature scaling behaviours 
for the full DC conductivity matrix that are determined by the scaling dimension,  $\Delta(k_1)$,
of the least irrelevant IR operator, thus 
generalising the results of \cite{Hartnoll:2012rj}. 
We find that
\begin{align}
\sigma \sim T^{2-2\Delta(k_1)},\quad \alpha \sim T^{2-2\Delta(k_1)},\quad \bar\kappa\sim T^{3-2\Delta(k_1)}\,.
\end{align}
with $\Delta(k_1)>1$. 
We see that while $\sigma$ and $\alpha$ will diverge
at $T=0$, $\bar\kappa$ will go to zero, if $1<\Delta(k_1)<3/2$, a constant, if $\Delta(k_1)=3/2$, and
diverge if $3/2<\Delta(k_1)$.
We also calculate $\kappa\equiv \bar\kappa-\alpha^2T/\sigma$, the thermal
conductivity at zero electric current, a quantity that is more readily measurable than $\bar \kappa$. Interestingly, we find
that for these $AdS_2\times\mathbb{R}^2$ ground states we have $\kappa\sim T$ independent of $\Delta(k_1)$.

The Q-lattice black hole solutions of \cite{Donos:2013eha,Donos:2014uba} also include black holes which at $T=0$ approach other 
IR solutions\footnote{Some of these IR solutions were also discussed in \cite{Gouteraux:2014hca}.} 
which break translation invariance and can be electrical insulators or metals depending on whether $\sigma$ is zero or non-zero respectively. Here we will find
that the insulating ground states and, somewhat surprisingly, 
also the metallic ground states, are both thermal insulators with $\kappa,\bar\kappa=0$ at $T=0$.
This suggests that the non-vanishing electrical conductivity of these metals can be thought of, loosely, 
as arising from the evolution of charged particle-hole pairs, possibly pair produced, in an electric field.

We also consider some other examples which involve linear massless axions which lead to momentum dissipation 
\cite{Azeyanagi:2009pr,Mateos:2011ix,Mateos:2011tv,Andrade:2013gsa} 
(see also \cite{Donos:2014uba,Gouteraux:2014hca}). 
Although these solutions share some similarities with the Q-lattice black holes of \cite{Donos:2013eha,Donos:2014uba}, 
they differ in several respects. While the Q-lattice 
is a periodic deformation of the UV physics by relevant operators of the CFT, the linear massless axion is a non-periodic deformation using
marginal operators. In the case that the linear axion black hole solutions approach $AdS_2\times\mathbb{R}^{D-2}$ in the far IR,
the linear axionic deformation is still present, in contrast to the Q-lattice where it vanishes as an irrelevant deformation.
We find that for the linear axion black holes of \cite{Mateos:2011ix,Mateos:2011tv,Andrade:2013gsa} that $\bar\kappa\to 0$ as $T\to 0$. 
More specifically for the $D=5$ black holes of \cite{Mateos:2011ix,Mateos:2011tv}, describing
anisotropic $N=4$ super Yang-Mills plasma,
we find that $\bar\kappa\sim T^{7/3}$.

The plan of the rest of the paper is as follows. In section \ref{sec2} we introduce the holographic models and black holes that we will consider in this paper.
In section \ref{sec3} we first calculate $\sigma$ and $\bar \alpha$ by deforming the black holes with an applied electric field but with no source for the heat current. This calculation highlights the utility of using a time-like killing vector in the bulk in order to obtain
the key expression for the heat current in terms of horizon data.  
We then complement this analysis by considering a thermal gradient in order to calculate $\bar\kappa$ and $\alpha$. Section \ref{sec4} illustrates our results with
some black hole solutions that have been constructed previously.
Section 5 concludes. We have three appendices.

\section{The holographic black holes}\label{sec2}

\subsection{The Holographic Models}
We will mostly focus on holographic models in $D=4$ spacetime dimensions which are dual to $d=3$ CFTs with a global $U(1)$ symmetry.
The $D=4$ fields include a metric and a gauge field, which are dual to the stress tensor and
the $U(1)$ current of the CFT, respectively. We will also include
a real scalar field, $\phi$, and two real  ``axion" fields, $\chi_i$, which are dual to additional scalar operators in the CFT.
The action is given by
\begin{align}\label{eq:aniso_model}
S=
\int d^4 x\sqrt{-g}\left[R-\frac{1}{2}\left[(\partial\phi)^2+\Phi_1(\phi)(\partial\chi_{1})^2+\Phi_2(\phi)(\partial\chi_{2})^2\right]-V(\phi)-\frac{Z(\phi)}{4}F^2\right]
\, ,
\end{align}
which involves four functions, $\Phi_i,V$ and $Z$, of the real scalar field $\phi$ and we demand $\Phi_i, Z\ge 0$.
Also, we have set $16\pi G=1$. We assume the model admits a unit radius $AdS_4$ vacuum
with $\phi=0$ (in particular $V(0)=-6$) and we shall choose $Z(0)=1$ for convenience.
The action is invariant under the global symmetries corresponding to shifts of the axion fields. 

For a holographic Q-lattice we are interested in the cases where the fields $\chi_i$ are necessarily periodic. 
These models arise when $\Phi_i\sim\phi^2$ near $\phi=0$.
For 
example, for a single axion (i.e. setting $\chi_2=0$), we could consider $\Phi_1=\phi^2$ and then $\phi,\chi_1$ are the norm
and phase of a complex scalar field. Furthermore we would choose the mass of this complex field, by choosing $V$, 
so that the complex field is dual to a relevant operator with dimension $\Delta<3$.
A deformation of the CFT by this complex operator with $\chi_1$ linear in a spatial direction would necessarily comprise a periodic deformation
and hence what we call a holographic lattice. Indeed decomposing the complex field into two real fields, reveals that
the construction has two real periodic lattices in the same spatial direction with a phase shift of $\pi/2$. 
This was precisely the construction of the anisotropic Q-lattices in \cite{Donos:2013eha}.  
Similarly the models with two $\chi_i$ can arise from two complex scalar fields with a $Z_2$ symmetry that equates their norms; these constructions lead
to isotropic Q-lattices as considered in \cite{Donos:2014uba}.
 
Our model also includes other types of black hole solutions where the $\chi_i$ are, instead, massless fields, and are dual to marginal operators with
$\Delta=3$. 
These models arise when $\Phi_i(0)\ne 0$.
For example, the case when $\Phi_i=1$ has been considered in \cite{Andrade:2013gsa}. Another case is for a single axion (i.e. setting $\chi_2=0$)
and $\Phi_1=e^{2\phi}$, corresponding to the axion and dilaton of string theory after performing a dimensional reduction
of type IIB supergravity on a five-dimensional Einstein space, and anisotropic black holes have been studied in
\cite{Mateos:2011ix,Mateos:2011tv,Cheng:2014qia}. In these cases, the linear axions do not give a periodic deformation of the CFT and hence should not be
considered as holographic lattices. Nevertheless, like the Q-lattices, they do incorporate momentum dissipation and have finite DC conductivities.

It is straightforward to generalise our results to other spacetime dimensions. For example, in $D=5$ we could add an additional
axion field $\chi_3$ along with a coupling $\Phi_3(\phi)$, in the obvious way. Our final result for the DC conductivities (see \eqref{findcres}) 
will be written in a way that is also valid for this case too.

\subsection{The black hole backgrounds}
The solutions that we shall consider all lie within the ansatz
\begin{align}\label{ansatz3}
&ds^{2}=-U\,dt^{2}+U^{-1}\,dr^{2}+e^{2V_1}dx_{1}^{2}+e^{2V_2}dx_{2}^{2},\nn
&A=a\,dt,\qquad \chi_{1}=k_1\,x_{1},\qquad \chi_{2}=k_2\,x_{2}\,,
\end{align}
where $U,V_i,a$ and $\phi$ are functions of $r$ only. 
In general the solutions are anisotropic, with $V_1\ne V_2$, but isotropic solutions with $V_1=V_2$ 
are possible when we can choose $k_1^2\Phi_1(\phi)=k_2^2\Phi_2(\phi)$.

We will assume that there is a regular event horizon at $r=r_+$ with the following expansions
\begin{align}\label{bgbhexp}
U&\sim4\pi T(r-r_+)+\dots,\qquad \quad V_i\sim V_{i+}+\dots,\nn
a&\sim a_+(r-r_+)+\dots,\qquad\quad \phi\sim \phi_++\dots\,,
\end{align}
where $T$ is the temperature of the black hole.
Below we will use ingoing Eddington-Finklestein coordinates $(v,r)$ where 
\begin{align}\label{efv}
v=t+\frac{1}{4\pi T}\ln(r-r_+)\,.
\end{align}

As $r\to\infty$, the location of the $AdS_4$ boundary, we assume that
\begin{align}\label{uvexpsss}
U&\sim r^2+\dots,\qquad\quad\,\,
e^{2V_i}\sim r^2+\dots\,,\nn
a&\sim\mu-q r^{-1}+...,\qquad
\phi\sim\lambda r^{\Delta-3}+\dots\,,
\end{align}
For the case of the Q-lattice, as discussed above, we would demand that $\Delta<3$ and $\lambda$ denotes the strength of the Q-lattice deformation (assuming a standard quantisation for the scalar). For the Q-lattice black holes the axions are periodic, $\chi_i=\chi_i+2\pi$, and these UV boundary conditions explicitly break the translation symmetry in a periodic manner.
The UV data specifying these black holes is given by $T/\mu$, $k_1/\mu$, $k_2/\mu$ and $\lambda/\mu^{3-\Delta}$.
For the case of massless linear axions, as discussed above, $\phi$ can also be massless or absent and the axions are not periodic.

It is useful to obtain a general expression for the electric charge 
of the black holes in terms of horizon data. 
The current density $J^a=(J^t,J^x,J^y)$ in the dual field theory has the form
\begin{align}\label{jdef}
J^a=
\sqrt{-g}Z(\phi)F^{ar}\,,
\end{align}
where the right hand side is evaluated at the boundary $r\to\infty$. The only non-zero component of the
equation of motion for the gauge-field is in the $t$-direction and can be written
$\sqrt{-g}\nabla_\mu(Z(\phi) F^{\mu t})=\partial_r(\sqrt{-g}Z(\phi)F^{r t})=0$. Thus we can write
\begin{align}
q\equiv J^t=e^{V_1+V_2}Z(\phi)a'\,.
\end{align}
where $q$ is the charge of the black hole and the right hand side can be evaluated at any value of $r$ including $r=r_+$. 
We note that the charge $q$ depends on the UV data of the Q-lattices including the temperature of the black hole.

\section{Calculating the DC conductivities}\label{sec3}

\subsection{Calculating $\sigma$ and $\bar \alpha$}
In this subsection we recall the derivation of \cite{Donos:2014uba}
for $\sigma$ and extend it to obtain $\bar\alpha$. 
Specifically, we switch on a constant electric field in the $x_1$ direction, with magnitude $E$, and no source for the heat current. 
For the black holes of interest, this will generate electric and heat currents just
in the $x_1$ direction, which we will label
$J\equiv J^{x_1}$ and $Q\equiv Q^{x_1}$, respectively. From \eqref{bigform} we see that expressions for $\sigma$ and $\bar\alpha$ (more precisely $\sigma^{x_1x_1}$ and 
$\bar\alpha^{x_1x_1}$) can be obtained once
we have obtained expressions for $J$ and $Q$. The following derivation will lead to expressions 
for $\sigma$ and $\bar\alpha$ in terms of horizon data. We can obtain the DC conductivity in the $x_2$ direction, when they are finite,
in an identical manner.

We consider the following small perturbation about the class of black hole solutions that we considered in the last section
\begin{align}\label{dcxan}
A_{x_1}&=-Et +{\delta a_{x_1}}(r)\,,\nn
g_{tx_1}&=\delta g_{tx_1}(r)\,,\nn
g_{rx_1}&=e^{{2V_1}}\delta h_{rx_1}(r)\,,\nn
\chi_1&=k_1x_1+\delta\chi_1(r)\,,
 \end{align}
which one can check is consistent with the linearised equations of motion. 

We first show that the linearised gauge-equations of motion imply that the current 
$J$ is constant and hence obtain an expression in terms of horizon data.
Specifically, the only non-trivial component of the gauge equation of motion is the $x_1$ component and it can be written in the form
$\partial_r(\sqrt{-g}Z(\phi)F^{r x_1})=0$. Using  \eqref{jdef} we deduce that $J=-\sqrt{-g}Z(\phi)F^{r x_1}$
is a constant. Explicitly we have
\begin{align}\label{jexp}
J=-e^{{V_2}-{V_1} }Z(\phi) U \delta a_{x_1}'-qe^{-2V_1} \delta g_{tx_1}\,,
\end{align}
where the right-hand side can be evaluated at any value of $r$, including at the black hole horizon at $r=r_+$.

We next consider the linearised Einstein equations.
We find one equation which we can algebraically  solve for $\delta h_{rx_1}$ giving
\begin{align}\label{gravcon}
\delta h_{rx_1}=\frac{Eqe^{-V_1-V_2}}{k_1^2\Phi_1(\phi)U}+\frac{\delta\chi_1'}{k_1}\,,
\end{align}
as well as the following second order ODE
\begin{align}\label{heq}
\delta g_{tx_1}''+(-{V_1}' +{V_2}')\delta g_{tx_1}'
- \left(2V_1'(V_1'+V_2')+2V_1''+ \frac{      e^{-2 {V_1}} {k_1}^2 {\Phi_1}(\phi) }{U}\right){\delta g_{tx_1}}\nn+e^{- {V_1}-{V_2}} q \delta{a_{x_1}'}  =0\,.
\end{align}
We also note that \eqref{gravcon} implies the equation of motion for $\delta\chi_1$.

Before studying these further we first discuss the boundary conditions that must be imposed on the linearised perturbation
at infinity and at the black hole horizon. Observe that
$\delta\chi_1$ only appears in \eqref{gravcon}; we will assume that $\delta\chi_1$ is analytic at the black hole event horizon and falls off 
sufficiently fast at infinity.
To ensure that the perturbation is regular as $r\to r_+$ i.e. at the black hole horizon, we need to switch from the coordinates $(t,r)$, which 
are ill-defined there,
and
employ Eddington-Finklestein coordinates $(v,r)$, with $v$ defined in \eqref{efv}. 
Specifically, in the $(t,r)$ coordinates, the gauge-field will be well defined if we demand that 
\begin{align}\label{bcax}
\delta{a_{x_1}}\sim-\frac{E}{4\pi T}\ln(r-r_+)+{\cal O}(r-r_+)\,,
\end{align}
since then the full gauge-field perturbation in \eqref{dcxan} 
has the regular expansion $A_{x_1}\sim -Ev +\dots$ in the Eddington-Finklestein coordinates.
Notice that near the horizon we have $\delta a_{x_1}'\sim-\frac{E}{U}+\dots$, a result that we will use below.
For the metric perturbation, we see from \eqref{gravcon} that $\delta h_{rx_1}$ is diverging; this can be remedied by demanding that
$\delta g_{tx_1}$ behaves as 
\begin{align}\label{hdpe}
\delta g_{tx_1}\sim -\frac{Eqe^{V_1-V_2}}{k_1^2\Phi_1(\phi)}|_{r=r_+}+{\cal O}(r-r_+)\,.
\end{align}
Notice that this behaviour for 
$\delta a_{x_1}'$ and $\delta g_{tx_1}$ is consistent with \eqref{heq}.

We next consider the behaviour as $r\to\infty$. From the expression for the gauge-field in \eqref{dcxan}
we see that we have a deformation of an electric field
in the $x_1$ direction with strength $E$. 
We also have the fall-off of $\delta a_{x_1}\sim Jr^{-1}$ which together
with \eqref{bcax} completely specifies $\delta a_{x_1}$.
Now, \eqref{heq} has two independent solutions, one of which behaves as 
$r^2$ and the other as $r^{-1}$; in order to have no additional deformations, associated with sources for the heat current, we demand that the coefficient of the former vanishes.
Observe that since we have also specified the boundary condition of
$\delta g_{tx_1}$ at the horizon this completely specifies the solution of \eqref{heq}. In addition, notice
from \eqref{gravcon} 
that
for suitable choices of $\delta\chi_1$ the fall-off of $\delta h_{rx_1}$ as $r\to\infty$
can be as weak as desired and with vanishing non-normalisable source for $\delta\chi_1$.

At this stage we have obtained a perturbation which is well defined in the bulk, including the horizon and we can use it to
obtain the DC conductivities $\sigma$ and $\bar\alpha$.
The DC electric conductivity is given by $\sigma=J/E$. To obtain our final result we now simply 
evaluate the right hand side of \eqref{jexp} at the black hole event horizon $r=r_+$. Doing so we obtain the expression found in \cite{Donos:2014uba}:
\begin{align}\label{dcresult}
\sigma=&
\left[e^{-{V_1}+{V_2}} Z(\phi)+\frac{e^{{V_1}+{V_2}}Z(\phi)^2(a')^2   }{{k_1}^2 {\Phi_1}(\phi)}\right]_{r=r_+}\,,\nn
=&
\left[ \frac{Z(\phi)s}{4\pi e^{2{V_1}}}+\frac{ 4\pi q^2}{{k_1}^2 {\Phi_1}(\phi)s}\right]_{r=r_+}\,,
\end{align}
where $s=4\pi e^{V_1+V_2}$ is the entropy density of the unperturbed black holes.

To obtain a similar expression for $\bar\alpha$ we need to obtain an expression for the heat current $Q$ analogous
to \eqref{jexp}. In essence (we also need to use the gauge equation of motion)  
we need to find a first integral of the equation of motion \eqref{heq}. 
While this can be guessed,
the underlying reason\footnote{Note that this would be obscure in the approach 
to calculate the DC conductivity used in 
\cite{Gouteraux:2014hca,Andrade:2013gsa,Davison:2013jba,Blake:2013bqa},
where a time dependence of the form $e^{-i\omega t}$ is
assumed for the electric field and then later the $\omega\to 0$ limit is taken.} can be clarified by introducing a two-form associated with the Killing vector field $\partial_t$.
We first observe that if $k$ is an arbitrary Killing vector, and hence $\nabla^\mu k^\nu=\nabla^{[\mu}k^{\nu]}$,
which satisfies $L_k F=L_k\phi=L_k\chi_i=0$, then we can define a two-form
$G$ by
 \begin{align}
 G^{\mu\nu}=\nabla^\mu k^\nu+
\frac{1}{2} Z(\phi)k^{[\mu}F^{\nu]\sigma}A_\sigma+\frac{1}{4}({\psi}-2\theta)Z(\phi)F^{\mu\nu}\,,
 \end{align}
 where $\psi$ and $\theta$ are defined by $L_k A=d\psi$ and $i_k F=d\theta$. It has the important 
 property that 
\begin{align}\label{divG}
 \nabla_\nu G^{\mu\nu}=-\frac{V}{2} k^\mu\,, 
 \end{align}
when the equations of motion are satisfied. A derivation is provided in appendix \ref{appkv}.

Focussing now on the Killing vector $k=\partial_t$, we consider the $x_1$ component of \eqref{divG} to deduce that
$\partial_r(\sqrt{-g}G^{rx_1})=0$
and hence that $\sqrt{-g}G^{x_1r}$ is a constant. We choose $\theta=-Ex_1 -a$ and $\psi=-Ex_1$, where we have fixed some free integration constants for convenience, to conclude that at {\it linearised order} we can write 
\begin{align}\label{pexp}
Q&\equiv 2\sqrt{-g}G^{rx_1}\,,\nn&=2\sqrt{-g}\nabla^{r} k^{x_1}+{a}\sqrt{-g}F^{rx_1}\,,\nn
&=e^{-V_1+V_2}U^2
\left(\frac{\delta g_{tx_1}}{U}\right)'
-a J\,,
\end{align}
where $Q$ is a constant. Since $Q$ is a constant we can evaluate the right hand side at any value of $r$. 
In particular, if we evaluate at the boundary $r\to\infty$, we find that $Q$ is indeed the heat 
current. The last term is simply $-\mu J$ while, as we explain in appendix B, the first term is 
$T^{tx_1}=r^5 \bar T^{tx_1}$ where  $\bar T^{\mu\nu}$ is the holographic stress tensor of
\cite{Balasubramanian:1999re} and hence
\begin{align}
Q=T^{tx_1}-\mu J\,.
\end{align}
We can also evaluate at the black hole horizon and at leading order in $(r-r_+)$ we deduce that
\begin{align}
&Q=-4\pi T e^{-V_1+V_2} \delta g_{tx_1}\,.
\end{align}
Then using \eqref{hdpe} 
we conclude that $\bar\alpha=Q/E$ can be expressed in terms of horizon data of the black holes as
\begin{align}
\bar\alpha&=\left[\frac{ 4\pi q}{{k_1}^2 {\Phi_1}(\phi)}\right]_{r=r_+}\,.
\end{align}

\subsection{Calculating $\alpha$ and $\bar\kappa$}
We now want to consider perturbations which have a source for the heat current.
This will allow us to obtain expressions for $\alpha$ and $\bar\kappa$.
To do this we will consider the following linearised 
perturbation about the black hole solutions that we considered in section 2:
\begin{align}\label{dcxan2}
A_{x_1}&=t\delta f_1(r)  +{\delta a_{x_1}}(r)\,,\nn
g_{tx_1}&=t\delta f_2(r)+\delta g_{tx_1}(r)\,,\nn
g_{rx_1}&=e^{{2V_1}}\delta h_{rx_1}(r)\,,\nn
\chi_1&=k_1x_1+\delta\chi_1(r)\,.
 \end{align}

We first consider the gauge equations of motion
\begin{align}\label{jexp2}
&\partial_r\left(\sqrt{-g}Z(\phi)F^{x_1r}\right)=0\,,
\end{align}
where
\begin{align}
&\sqrt{-g}Z(\phi)F^{x_1r}=-e^{{V_2}-{V_1} }Z(\phi)\left[\left(U \delta a_{x_1}'+a'\delta g_{tx_1}\right)
+t\left(U \delta f_{1}'+a'\delta f_{2}\right)\right]\,,
\end{align}
and we observe the explicit linear time dependence. 
We also find that one of the Einstein equations, combined with the gauge equations of motion, is equivalent to
the condition that $\tilde Q$ is independent of $r$ where
\begin{align}\label{qexp2}
\tilde Q&\equiv e^{-V_1+V_2}U^2\,\left[ 
\left(\frac{\delta g_{tx_1}}{U}\right)'  +t \left(\frac{\delta f_2}{U}\right)'
\right]-a  \sqrt{-g}Z(\phi)F^{x_1r}\,.
\end{align}
The remaining Einstein equation can be solved for $\delta h_{rx_1}$:
\begin{align}\label{gravcon2}
\delta h_{rx_1}=\frac{q\delta f_1e^{-V_1-V_2}}{k_1^2\Phi_1(\phi)U}+\frac{e^{2V_1}(e^{-2V_1}\delta f_2)'}{k_1^2 U\Phi_1}
+
\frac{\delta\chi_1'}{k_1}\,.
\end{align}

At this point we can observe that if we choose
\begin{align}\label{effsol}
\delta f_1&=-E+\hsce a(r)\,,\nn
\delta f_2&=-\hsce U(r)\,,
\end{align}
for constants $E,\hsce$ then all time dependence drops out of \eqref{jexp2}, \eqref{qexp2}.
Equivalently, this choice of $f_1,f_2$ solves the full linearised equations of motion.
We also note that when $\hsce=0$ then we have exactly the same set-up as in the last subsection.

Proceeding with this choice for $\delta f_1$ and $\delta f_2$ we find that \eqref{gravcon2} now implies 
the equation of motion for $\delta\chi_1$. We also find that defining
\begin{align}\label{jqexp3}
J&\equiv -e^{{V_2}-{V_1} }Z(\phi)\left(U \delta a_{x_1}'+a'\delta g_{tx_1}\right)\,,
\nn
Q&\equiv e^{-V_1+V_2}U^2\,
\left(U^{-1}\delta g_{tx}\right)' 
-aJ\,,
\end{align}
then both $J$ and $Q$ are constants. By evaluating at $r\to\infty$, it is clear that $J$ is the current in the $x_1$ direction. In appendix B
we will explain why $Q$ is the time-independent part of the heat current in the $x_1$ direction.

We now analyse regularity conditions at the black hole event horizon. For the gauge-field, as in the last subsection, 
we again need to impose that
\begin{align}\label{axre}
\delta a_{x_1}\sim -\frac{E}{4\pi T}\ln(r-r_+)+\dots\,,
\end{align}
where the dots refer to terms analytic in $(r-r_+)$. 
Again allowing $\delta\chi_1$ to be a constant on the horizon, we see from \eqref{gravcon2} that 
$\delta h_{rx}$ is diverging like $\sim U^{-1}$ at the horizon. By switching to Kruskal coordinates 
we can ensure regularity of the
linearised metric perturbation by choosing the behaviour of $\delta g_{tx}$ to behave near the horizon as
\begin{align}\label{gtxre}
\delta g_{tx}\sim Ue^{2 V_1}\delta h_{rx}|_{r=r_+}-\frac{\hsce U}{4\pi T}\ln(r-r_+)+\dots\,.
\end{align}
Observe (unlike in the last subsection) that this implies conditions on the leading and the sub-leading terms in the expansion abut $r=r_+$.
Remarkably \eqref{axre},\eqref{gtxre} are consistent with the second order equations for $\delta a_{x_1}$ and $\delta g_{tx_1}$ with 
first integrals as in \eqref{jqexp3}. 
Furthermore, we can
demand that the fall-off of $\delta g_{tx_1}(r)\sim r^{-1}$ and $\delta a_{x_1}(r)\sim Jr^{-1}$ at infinity, which is consistent with \eqref{jqexp3}.
To fully specify the perturbation, as in the last subsection, we suitably choose
$\delta\chi_1$ so that the fall-off of $\delta h_{rx_1}$ as $r\to\infty$
is as weak as desired, with vanishing non-normalisable source for $\delta\chi_1$.
Most importantly, 
we find that the expansions of the perturbation can be developed with the constants $J,Q$ given at the horizon by
\begin{align}\label{jqexp4}
J&=\left[E\left(e^{-V_1+V_2}Z(\phi)+\frac{e^{V_1+V_2}Z(\phi)^2(a')^2}{k_1^2\Phi(\phi)}\right)
+\hsce\frac{e^{V_1+V_2}Z(\phi)a'U'}{k_1^2\Phi(\phi)}\right]_{r=r_+}\,,\nn
Q&=\left[E\frac{e^{V_1+V_2}Z(\phi)a'U'}{k_1^2\Phi(\phi)}+\hsce\frac{e^{V_1+V_2}(U')^2}{k_1^2\Phi(\phi)}\right]_{r=r_+}\,,
\end{align}
and we note that we need to use the background equations of motion to get these expressions.
 
At this point we have obtained a linearised perturbation about the black holes solutions that
 is well defined on the black hole horizon and contains pieces that have a linear dependence in time.
 These time dependent pieces comprise the only holographic sources at the boundary at $r\to\infty$.
We have seen in the previous subsection that when $\hsce=0$ that $E$ parametrizes an electric field deformation.
As we explain in appendix B and C, upon setting $E=0$ we can deduce that $\hsce$ parameterises a time dependent
source for the heat current. Furthermore, using the results of appendix C, with independent $E,\hsce$ we can now calculate the full DC conductivity matrix, in the $x_1$ direction:
\begin{align}\label{findcres}
\sigma&=\frac{\partial}{\partial E}J
=\left[ \frac{Z(\phi)s}{4\pi e^{2{V_1}}}+\frac{ 4\pi q^2}{{k_1}^2 {\Phi_1}(\phi)s}\right]_{r=r_+}\,,\nn
\bar \alpha&=\frac{1}{T}\frac{\partial}{\partial E}Q=\left[\frac{ 4\pi q}{{k_1}^2 {\Phi_1}(\phi)}\right]_{r=r_+}\,,\nn
\alpha&=\frac{1}{T}\frac{\partial}{\partial \hsce}J=\left[\frac{ 4\pi q}{{k_1}^2 {\Phi_1}(\phi)}\right]_{r=r_+}\,,\nn
\bar\kappa&=\frac{1}{T}\frac{\partial}{\partial \hsce}Q=\left[\frac{4\pi sT} {{k_1}^2 {\Phi_1}(\phi)}\right]_{r=r_+}\,.
\end{align}
It is a satisfying check that we find $\alpha=\bar\alpha$ and hence a symmetric conductivity matrix.

\subsection{Comments}

Although we have carried out the derivation in $D=4$ space-time dimensions, the final expressions
\eqref{findcres} are also valid in other space-time dimensions. For example, in $D=5$ they are valid
when the model \eqref{eq:aniso_model} is generalised to have another axion $\chi_3$ with an associated
function $\Phi_3(\phi)$. The black hole ansatz \eqref{ansatz3} should also be generalised to have $\chi_3=k_3 x_3$, with isotropic black holes, with $V_1=V_2=V_3$, only possible when $k_1^2\Phi_1(\phi)=k_2^2\Phi_2(\phi)=k_3^2\Phi_3(\phi)$.

We next note that if one is interested in electrically neutral black holes then the results
\eqref{findcres} are also valid if we set $q=0$. For example, for the $D=4$ AdS-Schwarzschild black hole
we get $\sigma=Zs/(4\pi e^{2V_1})$, for constant $Z$, recovering the result of \cite{Iqbal:2008by}.
The fact that the first term in $\sigma$ is non-zero for neutral black holes with $k_1=0$ suggests that, loosely speaking,
it is associated with current flow arising from the evolution of charged particle-hole pairs, possibly
pair created, in an electric field. We should, however, bear in mind that there are no quasi-particles. Similarly, the second
term, as well as the expressions for $\alpha,\bar\kappa$, which diverge as $k_1\to 0$ can be then associated
with momentum dissipation. 
In fact, for general $q$, the first term in $\sigma$ has a simple interpretation as 
the conductivity in the absence of heat flows, i.e. with $Q=0$. Specifically, we have
\begin{align}
\left(\frac{J}{E}\right)_{Q=0}\equiv\sigma-\frac{\alpha\bar\alpha T}{
\bar\kappa}=\left[\frac{Z(\phi)s}{4\pi e^{2{V_1}}}\right]_{r=r_+}
\end{align}
where the first equality immediately follows from \eqref{bigform} and the second from \eqref{findcres}.

It is interesting to observe that for the class of black holes we are considering we always have the simple relation
\begin{align}\label{rat}
\frac{\bar\kappa}{\alpha}=\frac{Ts}{q}\,.
\end{align}
Next we recall that $\bar \kappa$ is the thermal conductivity at zero electric field. We can also define $\kappa$, the thermal
conductivity at zero electric current, a quantity that is more readily measurable. 
From \eqref{bigform} we deduce that $\kappa\equiv \bar\kappa-\alpha\bar\alpha T/\sigma$
and hence\begin{align}\label{kapex}
\kappa=\left[\frac{4\pi sT e^{2V_2}Z(\phi)}{q^2+k_1^2e^{2V_2}Z(\phi)\Phi_1(\phi)}\right]_{r=r_+}\,.
\end{align}
Unlike $\bar\kappa$, we see that $\kappa$ is well defined if we set $k_1^2\Phi_1(\phi)\to 0$.
Additional quantities of interest are the ratios of thermal conductivities to electric conductivities.  We find
\begin{align}\label{exl}
	\bar L&\equiv \frac{\bar\kappa}{\sigma T}=\left[\frac{s^2}{q^2+k_1^2e^{2V_2}Z(\phi)\Phi_1(\phi)}\right]_{r=r_+}\,,\nn
	L&\equiv \frac{\kappa}{\sigma T}=\left[\frac{k_1^2 e^{2V_2}s^2 Z(\phi)\Phi_1(\phi)}{\left(q^2+k_1^2e^{2V_2}Z(\phi)\Phi_1(\phi)\right)^2}\right]_{r=r_+}\,.
\end{align}
For Fermi liquids the ability of the quasi particles to transport heat is determined by their ability to transport charge
and $L$ is a constant\footnote{In fact when there is purely elastic scattering (either for very low $T$ for $T$ above the Debye temperature
where there is elastic phonon scattering) $L=\pi^2/3(k_B/e)^2$.}, as encapsulated in the Widemann-Franz law. 
Deviations from this behaviour is a possible indication
of strong interactions. It is also interesting to observe that $\bar L$ and $\kappa$ approach finite limits as $k_1\to 0$,
while $L$ approaches zero and $\bar\kappa$ diverges. 
We also note that we have the following bound
\begin{align}\label{newbd}
\bar L \le \frac{s^2}{q^2}\,,
\end{align}
for all of the black holes we have been considering.
Notice that this bound approaches saturation when the second term in the $\sigma$ in \eqref{findcres} dominates the first.

In the next subsection we will obtain the thermoelectric DC conductivities for various examples that have been
discussed in the literature. It is interesting to obtain the low-temperature scaling behaviours for the different ground
states that can arise. We find examples in which the two terms in $\sigma$ in \eqref{findcres} both scale in the same way. We
also find examples, which we might call ``pair evolution dominated", in which the first term in \eqref{findcres} dominates the second.
For both of these classes we find that $\kappa$ and $\bar\kappa$ scale in the same way. There is a third class of examples, which
we might call ``dissipation dominated", in which the second term in \eqref{findcres} dominates the first. In this case $\kappa$ and $\bar\kappa$
scale in different ways and we approach saturation of \eqref{newbd}. 
We will define the state to be metallic if it conducts at $T=0$ and electrically insulating if, instead, $\sigma= 0$. Note that for almost all
metallic states we will in fact have\footnote{Note that for arbitrarily small $T$, but $T\ne 0$, we have $\sigma$ is finite and hence 
so is $J$. Thus, $J$ can be made small by choosing
$E$ to be small enough and the set-up 
is suitable for extracting the linear response.} $\sigma\to\infty$ as $T\to 0$.

We end this section by commenting on the very high temperature behaviour of the conductivities.
When $T$ is much bigger than $\mu,k_i$ and the scale fixed by the lattice deformation strength $\lambda$, we can approximate the lattice black
holes by the $D=4 $ AdS-Schwarzschild black hole with $U=r^2-r_+^3/r$, $e^{2V_i}\sim r^2$ in \eqref{ansatz3}. With $r_+\propto T$ we conclude that
$s\sim T^{d-1}$, $q\sim T$.
Focussing on the case $\Phi_i\sim \phi^2$ as
$\phi\to 0$ (the case of periodic axions), we can solve the Laplacian for $\phi$ by first scaling the radial coordinate $r=r_+ \rho$. At leading order in $r_+$
there is no dependence on $r_+$ nor on $k$. We therefore have the solution with behaviour $\phi(\rho=1)$ is a constant independent of $r_+$, and, as
$\rho\to\infty$, $\phi(\rho)\sim A\rho^{\Delta-3}=r_+^{3-\Delta}A r^{\Delta-3}$, where $A$ is some constant. Now since we want the deformation at
infinity to be held fixed when we take the high temperature limit, we should rescale this solution by a factor of $r_+^{\Delta-3}$. Thus, at the horizon, $r=r_+$, we have $\phi\to T^{\Delta-3}$ and hence $\Phi_i(\phi)|_{r=r_+}\sim\phi^2|_{r=r_+}\sim T^{2(\Delta-3)}$. We thus conclude that the scaling in $\sigma$ is
dominated by the second term in \eqref{findcres} (dissipation dominated), and
\begin{align}
\sigma\sim T^{2(3-\Delta)},\quad \alpha\sim\sigma T,\quad \bar\kappa\sim\sigma T^3,\quad \kappa\sim T^3\,,
\end{align}
as $T\to\infty$.
This result is also valid for the case of linear axions with $\Delta=3$.
Note, in particular, that when $\Delta< 3$ we have a divergent $\sigma$ as $T\to \infty$. 
This can be contrasted with the behaviour of the optical conductivity $\sigma(\omega)$ which approaches a constant as $\omega\to\infty$. 
For the case of metals, in which the DC conductivity $\sigma$ is diverging at low temperatures, the fact that it also diverges at high temperatures when
$\Delta<3$ implies
that there will be a minimum conductivity at finite some temperature. This is reminiscent of the Mott-Ioffe-Regel bound \cite{Gunnarsson:2003zz,takmir}, 
but here, of course, we
have no quasi-particles.

\section{Examples}\label{sec4}

We illustrate our formulae for the DC conductivity using some $AdS$ black holes that exhibit momentum dissipation
which have been discussed in the literature. The simplest are the analytic isotropic 
solutions of \cite{Andrade:2013gsa} with
massless linear axions and so we present these first. We then discuss the anisotropic black holes with massless
linear axions of \cite{Mateos:2011ix,Mateos:2011tv} followed by the Q-lattice black holes of 
\cite{Donos:2013eha,Donos:2014uba}, all of which have been constructed numerically.

\subsection{Models with massless axions}
\subsubsection{Analytic isotropic solutions}\label{wanda}
A simple class of $D=4$ isotropic black hole solutions for \eqref{eq:aniso_model} with massless axions 
arises when 
\begin{align}\label{spvals}
\phi=0, \quad \Phi_i=1,\quad Z=1,\quad V=-6\,.
\end{align}
Indeed the solutions were constructed analytically in \cite{Bardoux:2012aw,Andrade:2013gsa} and are given by
\begin{align}
ds^2&=-f dt^2+\frac{dr^2}{f}+r^2(dx_1^2+dx_2^2)\,,\nn
A&=\mu(1-\frac{r_0}{r})\,,\nn
\chi_1&=k x_1,\qquad \chi_2=k x_2\,,
\end{align}
where $f=r^2-k^2/2-m_0/r+\mu^2r_0^2/4r^2$ with $m_0=r_0^3+r_0(\mu^2-2k^2)/4$.
The temperature of these black holes is related to $r_0$, the radial location of the black hole horizon, 
via
\begin{align}
r_0=\frac{2\pi }{3}\left(T+\sqrt{T^2+\frac{3(\mu^2+2 k^2)}{16\pi^2}}\right)\,.
\end{align}
At zero temperature the black hole approach $AdS_2\times\mathbb{R}^2$ in the far IR. 
Note that since $\Phi_i=1$ the fields $\chi_i$ do not have to be periodically identified for these black holes.

Using the formulae that we derived above we deduce that
\begin{align}
\sigma=1+\frac{\mu^2}{k^2},\qquad
\alpha=\frac{4\pi \mu}{k^2}r_0,\qquad
\bar\kappa=\frac{(4\pi)^2}{k^2}Tr_0^2\,,
\end{align}
as well as
\begin{align}
\bar L=\frac{(4\pi)^2}{(\mu^2+k^2)}r_0^2 \,,  \qquad L=\frac{(4\pi)^2k^2}{(\mu^2+k^2)^2}r_0^2\,,
\end{align}
which are also valid for the case $\mu=0$.
It is interesting to observe that while the electric conductivity $\sigma$ is finite at $T=0$, corresponding
to metallic behaviour, the thermal conductivity $\bar\kappa$ is going to zero. 
Notice that for these black holes the two terms in $\sigma$ in \eqref{findcres} both scale in the same way at low-temperatures.

\subsubsection{Anisotropic neutral black holes in $D=5$}
We next consider the anisotropic black holes in $D=5$ with a single massless axion field, linear in the
$x_1$ direction,
that were constructed in \cite{Mateos:2011ix,Mateos:2011tv} extending \cite{Azeyanagi:2009pr}.
These black holes are electrically neutral with no gauge field (i.e. $Z=0$) and have
\begin{align}\label{spvalss}
\Phi_i=e^{2\phi},\quad V=-12\,.
\end{align}
As $T\to 0$ the solutions approach Lifshitz solutions in the far IR that are
supported by the linear axion. The low temperature behaviour of
$\bar\kappa$ can be extracted from the finite temperature Lifshitz solutions which were found in \cite{Azeyanagi:2009pr}. The entropy density scales
with temperature as $s\sim T^{8/3}$ while the scalar scales as $e^{2\phi}\sim T^{4/3}$. Hence we conclude that as $T\to 0$, in the direction $x_1$ of the linear axion,
\begin{align}
\bar\kappa\sim T^{7/3}\,.
\end{align}
Thus these black holes are dual to ground states which are thermally insulating in the direction of the linear axion. In the
other spatial directions, $x_2$ and $x_3$, 
they are ideal thermal conductors with infinite $\bar\kappa$.

For the black holes constructed in \cite{Mateos:2011ix,Mateos:2011tv} one can also 
ask about the electric conductivity. It is natural to consider this question within the context
of the bosonic part of $D=5$ minimal supergravity coupled to  the axion-dilaton considered
in \cite{Mateos:2011ix,Mateos:2011tv}. By extending the arguments in \cite{Gauntlett:2007ma} 
one can
show that this theory arises as a consistent Kaluza-Klein truncation of $D=10$ string theory
on an internal manifold $M_5$ associated with any supersymmetric $AdS_5\times M_5$ solution
of the $D=10$ supergravity theory. Now minimal supergravity has a kinetic term for the Maxwell
field combined with a Chern-Simons term. 
However, for the calculation of the DC conductivity for the class of black holes considered in \cite{Mateos:2011ix,Mateos:2011tv}, the Chern-Simons term
plays no role and we can still use the formula for $\sigma$ 
given in \eqref{findcres} with $q=0$ and $Z=1$. By carrying out a similar analysis as above
we find that the low-temperature scaling of the electrical conductivity in the $x_1$ direction
is given by $\sigma \sim T^{4/3}$ while in the $x_2$ and $x_3$ directions it is given
by $\sigma \sim T^{2/3}$.

\subsection{Holographic Q-lattices}
Various holographic Q-lattice black hole solutions were constructed in 
\cite{Donos:2013eha,Donos:2014uba}, both isotropic and
anisotropic. Some of these approach $AdS_2\times\mathbb{R}^2$ in the IR at $T=0$ while others approach
new IR ground states which were independently found in \cite{Gouteraux:2014hca}. 

\subsubsection{Q-lattice black holes with $AdS_2\times\mathbb{R}^2$ in the IR at $T=0$}\label{memmat}
Consider Q-lattice deformations which at zero temperature approach electrically charged $AdS_2\times\mathbb{R}^2$ solutions
in the near horizon limit. As $T\to 0$, these black holes will have $e^{V_i}|_{r=r_+}, Z|_{r=r_+},q$ and $s$ all
approaching non-zero constant values. On the other hand, for the holographic Q-lattice black holes 
(unlike the solutions in section \ref{wanda} above)
as $T\to 0$ we have $\Phi_1(\phi)\to 0$ near the horizon. More precisely, we have
$\Phi_1(\phi)\sim T^{2\Delta(k_1)-2}$
where $\Delta(k_1)>1$ is the dimension of
the irrelevant operator arising from perturbations of the scalar field $\phi$ about the 
$AdS_2\times\mathbb{R}^2$ background\footnote{Since there is a renormalisation of length scales from the UV to the IR, the $k_1$
appearing in $\Delta(k_1)$ is not the UV lattice momentum $k_1$.}. Explicit details of this calculation are presented in the examples of 
\cite{Donos:2013eha,Donos:2014uba}. Thus we immediately deduce that the low temperature scaling of
$\sigma$ is dominated by the second term in \eqref{findcres} (i.e. is ``dissipation dominated")
and we have
\begin{align}
\sigma \sim T^{2-2\Delta(k_1)},\quad \alpha \sim T^{2-2\Delta(k_1)},\quad \bar\kappa\sim T^{3-2\Delta(k_1)}\,.
\end{align}
The result for $\sigma$ agrees with the arguments of \cite{Hartnoll:2012rj}, using the memory matrix formalism,
which are valid for small lattice perturbations about translationally invariant IR ground states. 
It is interesting to observe that while $\sigma$ and $\alpha$ will diverge
at $T=0$, $\bar\kappa$ will go to zero, if $1<\Delta(k_1)<3/2$, a constant, if $\Delta(k_1)=3/2$, and
diverge if $3/2<\Delta(k_1)$.
We also find that as $T\to 0$ we approach a saturation of the bound on $\bar L$ given in \eqref{newbd}:
\begin{align}
\bar L \to \frac{s^2}{q^2}\,.
\end{align}
We also find the following low-temperature scaling behaviours
\begin{align}
\kappa\sim T,\qquad L\sim T^{2\Delta(k_1)-2}\,.
\end{align}
Notice that $\bar\kappa$ and $\kappa$ scale in different ways, as do $\bar L$ and $L$.

Finally, we note that reference \cite{Mahajan:2013cja} used the memory matrix formalism
to argue that these holographic black holes will have, approximately,
\begin{align}\label{bdsstext}
\bar L\sim\frac{1}{T^2}\frac{\chi^2_{QP}}{\chi^2_{JP}}\,,
\end{align}
where, in the notation of \cite{Mahajan:2013cja}, $\chi$ are static susceptibilities involving the operators for the total momentum $P$, electric current $J$ and heat current $Q$.
Similarly, using the results and notation of \cite{Mahajan:2013cja} we
conclude that these holographic black holes have, approximately,
\begin{align}\label{rat2}
\frac{\bar\kappa}{\alpha}\sim \frac{\chi_{QP}}{\chi_{JP}}\,.
\end{align}

\subsubsection{The Q-lattice black holes of \cite{Donos:2014uba}}
Various holographic Q-lattice black hole solutions were discussed in \cite{Donos:2014uba}. 
The most explicit constructions, presented in sections 2 and 3 of \cite{Donos:2014uba}, 
involved anisotropic lattices with a single axion field i.e. $\Phi_2=\chi_2=0$
and specific choices for $\Phi_1, V$ and $Z$ that involved a free parameter $\gamma$ with $-1<\gamma$.
Depending on the value of $\gamma$ it was shown that there can be metal-insulator as well as metal-metal transitions driven
by the strength of the holographic lattice deformation. 

In some cases at $T=0$ the black holes approach $AdS_2\times\mathbb{R}^2$ in the far IR, with non-vanishing entropy
density. In other cases the $T=0$ black holes approach new ground states, breaking translation invariance, which were presented
in section 2 of \cite{Donos:2014uba} (and also in \cite{Gouteraux:2014hca}). By analysing the small temperature behaviour of these ground states
by heating them up (i.e. by constructing small black hole solutions) some calculation reveals 
the following scaling behaviours as $T/\mu\to 0$:
 \begin{align}
\sigma\sim T^{\frac{(1+\gamma)(3-\gamma)}{9+2\gamma+\gamma^2}},\qquad
\alpha\sim T^{\frac{4(1+\gamma)}{9+2\gamma+\gamma^2}},\qquad
\kappa,\bar\kappa\sim T^{\frac{2(7+4\gamma+\gamma^2)}{9+2\gamma+\gamma^2}},\qquad
L,\bar L\sim T^{\frac{2(1+\gamma)^2}{9+2\gamma+\gamma^2}}\,.
\end{align}
Note that for $\sigma$,
both terms in \eqref{findcres} scale in the same way. In addition $q$ scales like $T^0$ as does
the other term appearing in the denominator of \eqref{kapex} and \eqref{exl}. By considering $\sigma$ we deduce that
for $-1<\gamma<3$ the ground states are insulators, while if $3\le \gamma$ they are metals.
Observe that for $-1<\gamma$ the exponents in $\alpha$ and
$\bar L, L$ are greater than zero, while the exponent for
$\bar\kappa, \kappa$ is greater than one. The fact that $\bar\kappa,\kappa\to 0$ at $T=0$ says that
a heat gradient does not give rise to transport. On the other hand if $3\le \gamma$ there is
transport of charge. This indicates that the latter transport can be loosely thought of as due to
the evolution of charged particle/anti-particle pairs (even though these metals are not ``pair creation dominated" as defined below \eqref{newbd}).

In section 4 of \cite{Donos:2014uba} (and also in \cite{Gouteraux:2014hca})
a different class of metallic and insulating ground state
solutions, isotropic in the spatial directions, were constructed which depended on three constants $c,\alpha$ and $\gamma$. 
For reasons that were explained in \cite{Donos:2014uba}, it was natural to focus on the range
\begin{align}\label{chgeconditions}
2\le \alpha<\sqrt{4+c},\qquad \gamma\ge \alpha-4\,.
\end{align}
The far IR of the ground states are electrically neutral solutions to the equations of motion with vanishing gauge-field, but the conditions \eqref{chgeconditions} imply that one can 
shoot out with an irrelevant or marginal operator to match on to the UV with $\mu\ne 0$.
By analysing the small black hole solutions, we find the following low temperature behaviour.
We find that $q$ is again independent of $T$ and that 
\begin{align}
\sigma\sim T^{-\frac{2(\alpha-2)\gamma}{4+c-\alpha^2}},\qquad
\alpha\sim T^{\frac{4(\alpha-2)}{4+c-\alpha^2}},\qquad
\kappa, \bar\kappa\sim T^\frac{4+c-4\alpha+\alpha^2}{4+c-\alpha^2},\qquad
L,\bar L\sim T^{\frac{2(\alpha-2)(\gamma+\alpha)}{4+c-\alpha^2}}\,.
\end{align}
The first term appearing in $\sigma$ in \eqref{findcres} now dominates the second term (i.e. they are ``pair-evolution" dominated). 
Similarly the second
term in the denominators of
\eqref{kapex}, \eqref{exl} dominate the first. 
By considering $\sigma$ we deduce that
for $\gamma\ge 0$ or $\alpha=2$ the ground states are electrical insulators, while if $3\le \gamma$ they are metals.
When $\alpha\ne 2$ we always have $\kappa, \bar\kappa\to 0$ at $T=0$ which again indicates that
for the metallic states the transport in an electric field might be viewed as 
arising from evolution of charged particle-hole pairs.
A special case is when $\alpha=2$ where both $\sigma$ and $\kappa, \bar\kappa$ go to a constant at $T=0$. However, it
is not yet clear if solutions which asymptote to $AdS_4$ exist when $\alpha=2$.

\section{Final Comments}
The main result of this paper is an expression for the thermoelectric DC conductivity matrix for a class
of asymptotically AdS black holes in terms of black hole horizon data. 
To achieve this we introduced sources for the electric and heat currents
that are 
linear in time. The full linearised perturbation also contains a time independent piece and we showed that
it was possible to obtain expressions for the time independent pieces of the electric and heat currents as total derivatives in the radial
direction which could then be written in terms of horizon data. For the electric current this step arises directly from the 
gauge equations of motion while for the heat current we saw that it arises 
from the existence of time-like Killing vectors for the time independent
perturbation. The final step to obtain the conductivity was to ensure regularity of the perturbation at the black hole horizon.

We obtained some general conditions on the conductivity including \eqref{ratintro}. For small lattice deformations, for which the memory matrix applies,
these can be recast as \eqref{newbdintro}.
Using our new results we obtained the thermoelectric DC conductivity for several explicit examples finding some interesting results. 
For example, the zero temperature ground states can be dissipation dominated, when the second term in $\sigma$ in \eqref{findcres} dominates the first,
and then $\bar\kappa$ and $\kappa$ scale in different ways. These include examples in which the ground states approach in the IR, $AdS_2\times\mathbb{R}^2$ deformed by irrelevant operators. We also found that the isotropic and the anisotropic ground states found in \cite{Donos:2014uba}
which break translation invariance, are all thermal insulators despite the fact that they can be electrically insulating or conducting. The isotropic ground states
are pair-evolution dominated, with the first term in $\sigma$ in \eqref{findcres} dominating the second, while for the anisotropic ground states the two terms
are equally important at low temperatures.

The black holes we have considered are homogeneous in the holographic directions. While they can be spatially
both isotropic and anisotropic, the conductivity matrix is diagonal and furthermore the models have $\alpha=\bar\alpha$ because of the underlying time-reversal invariance. However, our approach can be extended to more general set-ups
as we will explain in \cite{Donos:2014yya}.

We end by briefly commenting on two papers that appeared very recently. In \cite{Mefford:2014gia} black holes involving a scalar field
and interpolating between two $AdS_4$ geometries were constructed. These black holes fall within the class of solutions considered here and in 
\cite{Donos:2014uba} and have zero charge density, $q=0$. In particular, 
the formula for the DC conductivity given in \cite{Donos:2014uba}, which is valid in the $q=0$ limit, implies that
$\sigma\sim Z(\phi)$.
The model of \cite{Mefford:2014gia} is arranged so that $Z(\phi)\to 0$ in the far IR, leading to a vanishing electrical
conductivity, with a power law behaviour in $T$. 
Clearly the precise power will depend not only on the IR scaling dimension of the operator dual to $\phi$, but also
on the choice of $Z$.
Note, also, that
this model has no mechanism for momentum dissipation and hence there is a delta function in the thermal conductivity. It is also worth noting that since
$Z\to0$ in the IR, it will not be possible to add charge to these systems and furthermore quantum corrections will be important.

The second paper, \cite{Amoretti:2014zha}, calculates the thermoelectric response in the context of massive gravity, developing
the work of \cite{Davison:2013jba,Blake:2013bqa}.
The optical conductivities were calculated numerically, and from this it was possible to extrapolate some behaviour of the DC conductivity.
Our results here indicate that it should also be possible to obtain analytic results for the DC thermoelectric conductivities for the models considered in \cite{Amoretti:2014zha}.

\section*{Acknowledgements}
We thank Mike Blake and David Tong for helpful discussions. 
The work is supported by STFC grant ST/J0003533/1, EPSRC programme grant EP/K034456/1 and also by the European Research Council under the European Union's Seventh Framework Programme (FP7/2007-2013), ERC Grant agreements STG 279943 and ADG 339140.

\appendix
 
 \section{Killing vector identity}\label{appkv}
Suppose that $k$ is a Killing vector. 
We have, by definition, $\nabla_{(\mu} k_{\nu)}=0$ and hence $\nabla^\mu k^\nu=\nabla^{[\mu} k^{\nu]}$.
In addition we will suppose that we have $L_k F=L_k\phi=L_k\chi_i=0$. We will work in $D$ spacetime dimensions.
 We first observe that using the Bianchi identity
 we have $ (i_k d+d i_k)F=d(i_k F)=0$ and hence
 \begin{align}
k^\mu F_{\mu\nu}=\nabla_\nu \theta\,,
 \end{align}
 for some function $\theta$.
Using this and the equation of motion for the gauge-field, $\nabla_\mu(Z(\phi)F^{\mu\nu})=0$, we deduce that
 \begin{align}
k^\rho Z(\phi)F^2_{\rho\mu}=\nabla_\rho(\theta Z(\phi)F_{\mu}{}^\rho)\,.
\end{align}
By writing
$(L_k F)_{\mu\nu}=k^\mu\nabla_\mu F_{\nu \rho}+\nabla_\nu k^\mu F_{\mu\rho}+\nabla_\rho k^\mu F_{\nu\mu}=0$ we can similarly show
\begin{align}
k^\mu Z(\phi)F^2=4\nabla_\rho(Z(\phi)k^{[\mu}F^{\rho]\nu}A_\nu)+2\nabla_\rho(Z(\phi)F^{\mu\rho}\psi)\,,
\end{align}
where $\psi$ is defined via $L_kA=d\psi$.
 
We can now calculate
 \begin{align}
 \nabla_\mu(\nabla^\nu k^\mu)&=R^{\nu}{}_\mu k^\mu\,,\nn
 &=\frac{V}{D-2}k^\nu+\frac{1}{2}k^\mu Z(\phi)F^2_{\mu}{}^{\nu}-\frac{1}{4(D-2)}k^\nu Z(\phi)F^2\,,
 \end{align}
 where we used the Einstein equations and $L_k\chi_i=0$ to get the second line.
We thus conclude that when the equations of motion are satisfied we have
 \begin{align}
 \nabla_\mu G^{\mu\nu}=-\frac{V}{D-2} k^\nu\,,
 \end{align}
 where the two-form $G$ is given by
 \begin{align}
 G^{\mu\nu}=\nabla^\mu k^\nu+
\frac{1}{D-2} Z(\phi)k^{[\mu}F^{\nu]\sigma}A_\sigma+\frac{1}{2(D-2)}({\psi}-(D-2)\theta)Z(\phi)F^{\mu\nu}\,.
 \end{align}
 
 Note that we have
 \begin{align}
 \partial_\mu(\sqrt{-g}G^{\mu\nu})=-\frac{V}{D-2}\sqrt{-g} k^\nu\,,
 \end{align}
 and if we have $k=\partial_t$ then we have the right hand side is zero unless $\nu=t$. 

\section{Holographic renormalisation and the heat current}
The action \eqref{eq:aniso_model} should be supplemented with suitable boundary terms. For illustration we assume that
we are considering a holographic Q-lattice and in the $AdS_4$ vacuum the field $\phi$ is dual to a relevant operator with 
dimension $\Delta=2$. Then, for the solutions of interest,
we should use
\begin{align}
S_{ct}=\int d^3x\sqrt{-\gamma}(2K-4-\frac{1}{2}\phi^2+\dots)\,,
\end{align}
where $K$ is the trace of the extrinsic curvature and we have neglected additional terms involving the Ricci scalar which we don't need. 
Following
\cite{Balasubramanian:1999re}, the stress tensor and the current are given by
\begin{align}\label{genpertap}
\frac{1}{2}\bar T^{\mu\nu}&=-\left[K^{\mu\nu}-K\gamma^{\mu\nu}+\frac{1}{2}(4+\frac{1}{2}\phi^2)\gamma^{\mu\nu}+\dots\right]\,,\nn
\bar J^\nu&=-n_\mu F^{\mu\nu} Z(\phi)\,,
\end{align}
where the neglected term in the first line involves the components of the Einstein tensor for the boundary metric which won't contribute, and the right hand side of both lines are evaluated at the boundary $r\to\infty$. Note that in our conventions, on-shell we have
\begin{align}
\delta S=\int d^3x\sqrt{-\gamma}\left(\frac{1}{2}\bar T^{\mu\nu}\delta\gamma_{\mu\nu}+\bar J^\mu\delta A_\mu\right)\,.
\end{align}

For the black hole backgrounds we have $n_\mu=(0,1/U^{1/2},0,0)$ and we find that
\begin{align}
\bar T^{tt}&=U^{-1}\left(4+\frac{1}{2}\phi^2-2U^{1/2}(V_1'+V_2')\right)\,,\nn
\bar T^{xx}&=e^{-2V_1}\left(  -4-\frac{1}{2}\phi^2+U^{-1/2}U'+2U^{1/2}V_2'  \right)\,,\nn
\bar T^{yy}&=e^{-2V_2}\left(  -4-\frac{1}{2}\phi^2+U^{-1/2}U'+2U^{1/2}V_1'  \right)\,,\nn
\bar J^t&=\frac{Z(\phi)}{U^{1/2}}a'\,.
\end{align}
As $r\to\infty$ we have $\bar T^{\mu\nu}\sim r^{-5}$ and $\bar J^\mu \sim r^{-3}$
so it is convenient to define
\begin{align}\label{endtwo}
T^{\mu\nu}=r^5\bar T^{\mu\nu}\,,\qquad
J^\mu=r^3 \bar J^{\mu}\,.
\end{align}
Note that this is consistent with the definition of $J^{\mu}$ given in \eqref{jdef}.

We next consider the perturbation \eqref{dcxan2} about the background, but with
a general $g_{tx_1}(t,r)$ for the moment, finding
\begin{align}\label{bingop}
\bar T^{tx_1}
&=\frac{e^{-2V_1}}{U^{1/2}}\left( g_{tx_1}(t,r)[2V_2'-U^{-1/2}(4+\frac{\phi^2}{2})]+\partial_r g_{tx_1}(t,r)       \right)\,,
\end{align}
It will be convenient, shortly, to note that\footnote{Observe, in passing, the similarity of the left hand side with equation (15) of \cite{Balasubramanian:1999re}.}
\begin{align}\label{convexp}
U^{1/2}e^{V_1+V_2}\left(U \bar T^{tx_1}-g_{tx_1}(t,r)\bar T^{x_1x_1}\right)=e^{-V_1+V_2}U^2\partial_r\left(\frac{g_{tx_1}(t,r)}{U}\right)\,.
\end{align}

We now consider the particular linearised time-dependence for the perturbation given in \eqref{dcxan2}. The most important pieces of the
perturbation of relevance here are
\begin{align}\label{pertap}
A_{x_1}&=-tE+t\hsce a(r)  +{\delta a_{x_1}}(r)\,,\nn
g_{tx_1}(t,r)&=-t\hsce U+\delta g_{tx_1}(r)\,,
\end{align}
with $\delta a_{x_1}, \delta g_{tx_1}(r)\sim r^{-1}$ as $r\to \infty$.
We are interested in obtaining the expectation values for the $J^{x_1}$ and $T^{tx_1}$. Now in the text we showed
the time-dependent sources given in \eqref{dcxan2} give rise to a time-independent expression for $J^{x_1}$ given in 
\eqref{jqexp3}. However we obtain a time-dependent source for $T^{tx_1}$. Explicitly, from \eqref{bingop}
we immediately obtain 
\begin{align}
\bar T^{tx_1}
&={e^{-2V_1}U^{-1/2}}\left( \delta g_{tx_1}(r)[2V_2'-U^{-1/2}(4+\frac{\phi^2}{2})]+
\partial_r \delta g_{tx_1}(r)       \right)-\hsce t \bar T^{x_1x_1}\,,\nn
&\equiv \bar T^{tx_1}_0-\hsce t \bar T^{x_1x_1}\,.
\end{align}
Returning now to \eqref{convexp} and 
substituting in \eqref{pertap} we find that all of the time dependence drops out and hence we can conclude that
\begin{align}
U^{1/2}e^{V_1+V_2}\left(U \bar T^{tx_1}_0-\delta g_{tx_1}(r)\bar T^{x_1x_1}\right)=e^{-V_1+V_2}U^2\partial_r\left(U^{-1}\delta g_{tx_1}(r)\right)\,.
\end{align}
Evaluating both sides at $r\to\infty$ we deduce that
\begin{align}\label{btwelve}
r^5\bar T^{tx_1}_0=e^{-V_1+V_2}U^2\partial_r\left(U^{-1}\delta g_{tx_1}(r)\right)|_{r\to\infty}\,.
\end{align}
Recalling the expression for $Q$ given in \eqref{jqexp3}, we deduce that
\begin{align}
T^{tx_1}-\mu J^{x_1}=Q-\hsce t T^{x_1x_1}
\end{align}

We now ask how this time dependent response fits our expectations. 
The sources in our perturbation \eqref{pertap} are time dependent. After substituting into
\eqref{genpertap} we deduce that
\begin{align}
\delta S=\int d^3 x\left[  (T^{tx_1}-\mu J^{x_1})(-\hsce t)+J^{x_1}(-Et)\right]
\end{align}
In particular, this shows that $-\hsce$ is parametrizing a time dependent source for the operator $(T^{tx_1}-\mu J^{x_1})$.
In appendix \ref{linsce} we show that sources which are linear in time give a response, captured in the expectation value
of the operators, that contains, in general, a piece that is linear in time and a time-independent piece - see equation 
\eqref{finrst}. The time dependent piece is determined by the associated Greens function matrix at zero frequency and defines a static susceptibility. 
Since the only
linear time dependence appears in \eqref{btwelve} we deduce that the only non-zero component of this matrix is the
$Q^{x_1}$, $Q^{x_1}$ component, where $Q^{x_1}=T^{tx_1}-\mu J^{x_1}$, with $\tilde G_{Q^{x_1} Q^{x_1}}(0)=T^{x_1x_1}$.

Our principle interest is the DC conductivity, as defined in terms of the spectral weight in \eqref{intstepc2}. From \eqref{finrst} we see that this is captured by the time-independent pieces
of the expectation values. Using the results of this appendix and those in the text, then leads to the prescription that we employed
in \eqref{findcres}.

\subsection{A complementary point of view}
Let us consider the following coordinate transformation on the boundary:
\begin{align}
t&=\bar t -\hsce \bar t \bar x_1\,,\nn
x_1&=\bar x_1\,.
\end{align}
At {\it linearised order} we find that in these co-ordinates we have
\begin{align}\label{endone}
T^{\bar t\bar t}&=(1+2\hsce \bar x_1)T^{tt}\,,\quad T^{\bar x_1\bar x_1}=T^{x_1x_1}\,,\quad
T^{\bar t\bar x_1}
= T^{tx_1}_0\,,\nn
J^{\bar t}&=(1+\hsce \bar x_1)J^{t}\,,\quad J^{\bar x_1}=J^{x_1}\,,
\end{align}
and, in particular, the time-dependence has dropped out of the expectation values.

Let us now see how this coordinate transformation effects the asymptotic behaviour of the bulk fields at $r\to\infty$:
\begin{align}\label{ncs}
ds^2&=-U(1-2\hsce \bar x_1)d\bar t^2+r^2 d\bar x_1^2\dots\,,\nn
A&=a(1-\hsce \bar x_1)d\bar t -\bar t E d\bar x_1
+\dots\,.
\end{align}
When $E=0$ we can identify $-\hsce$ as a source for a static thermal gradient in the $\bar x_1$ direction, $(\partial_{\bar x_1} T)/T$,
with no source for the electric field. This again leads to the prescription that we employed
in \eqref{findcres}.

 \section{DC transport from linear sources in time}\label{linsce}
We consider a set of sources $s_{A}\left(t\right)$ associated with a set of operators $\phi_{A}$. At the level of linear response
we have 
\begin{align}
\left<\phi\left(t\right) \right>_{B}=\int dt^{\prime}\,G_{BA}\left(t-t^{\prime}\right)\,s_{A}\left(t'\right)\,,
\end{align}
where $G$ is the associated retarded Green's function and
we will define the Fourier transform as
\begin{align}
G_{BA}\left(t\right)=\frac{1}{2\pi}\,\int d\omega \, \tilde{G}_{BA}\left(\omega\right)\,e^{-i\,\omega\,t}.
\end{align}
We now examine the implications of sources linear in time $s_{A}\left(t\right)=c_{A}\,t$ leading to
\begin{align}
\left<\phi\left(t\right) \right>_{B}=\frac{1}{2\pi}\,\int dt^{\prime}\,d\omega\, e^{-i\,\omega\,\left(t-t^{\prime}\right)}\,t^{\prime}\,\tilde{G}_{BA}\left(\omega\right)\,c_{A}\,.
\end{align}
Using
\begin{align}
\int\,dt^{\prime}\,e^{i\,\omega\,t^{\prime}}\,t^{\prime}=-2\pi\,i\,\delta^{\prime}\left(\omega\right)\,,
\end{align}
we obtain
\begin{align}\label{intstepc}
\left<\phi\left(t\right) \right>_{B}&=i\,\partial_{\omega}\left(e^{-i\,\omega\,t}\,\tilde{G}_{BA}\left(\omega\right) \right)_{\omega=0} \,c_{A}\nn
&=\left( t\,\tilde{G}_{BA}\left(0\right)+i\,\tilde{G}^{\prime}_{BA}\left(0\right) \right)\,c_{A}.
\end{align}
We next define
the conductivity matrix $\sigma_{BA}$ as the zero-frequency limit of the spectral weight:
\begin{align}\label{intstepc2}
\sigma_{BA}=\lim_{\omega\rightarrow 0} \mathrm{Im} \frac{\tilde{G}_{BA}\left(\omega\right)}{\omega}\,.
\end{align}
Using the fact that the real and the imaginary parts of the Green's function are even and odd functions of $\omega$, respectively, we can rewrite
\eqref{intstepc} as 
\begin{align}\label{finrst}
\left<\phi\left(t\right) \right>_{B}=\left( t\,\tilde{G}_{BA}\left(0\right)-\sigma_{BA} \right)\,c_{A}.
\end{align}

\bibliographystyle{utphys}
\bibliography{helical}{}
\end{document}